\documentclass[reprint,showkeys,superscriptaddress,amsmath,amsfonts]{revtex4-1}
\newcommand\textsubscript[1]
  {\ensuremath{_{\mbox{\scriptsize #1}}}}
\usepackage{lmodern}
\usepackage{textgreek}
\usepackage{natbib}
\usepackage{booktabs}
\usepackage{graphics}
\usepackage[unicode=true]{hyperref}
\hypersetup{breaklinks=true,
            bookmarks=true,
            pdfauthor={},
            pdftitle={Spintronic properties and stability of the half-Heusler alloys LiMnZ (Z=N, P, Si)},
            colorlinks=true,
            citecolor=blue,
            urlcolor=blue,
            linkcolor=magenta,
            pdfborder={0 0 0}}
\begin{document}
\title{Spintronic properties and stability of the half-Heusler alloys LiMnZ
(Z=N, P, Si)}
\author{L. Damewood}
\affiliation{Department of Physics, University of California, Davis, CA 95616 USA}
\email{damewood@physics.ucdavis.edu}
\author{B. Busemeyer}
\affiliation{Department of Physics, University of Illinois, Urbana, IL 61801 USA}

\author{M. Shaughnessy}
\affiliation{RTBiQ, Inc., San Francisco, CA 94121 USA}

\author{C. Y. Fong}
\affiliation{Department of Physics, University of California, Davis, CA 95616 USA}

\author{L. H. Yang}
\affiliation{Lawrence Livermore National Laboratory, Livermore, CA 94551 USA}

\author{C. Felser}
\affiliation{Institut f\"{u}r Anorganische Chemie und Analytische Chemie, Johannes
Gutenberg-Universit\"{a}t Mainz, 55099 Mainz, Germany}

\date{\today}
\begin{abstract}
Li-based half-Heusler alloys have attracted much attention due to their
potential applications in optoelectronics and because they carry the
possibility of exhibiting large magnetic moments for spintronic
applications. Due to their similarities to metastable zinc blende
half-metals, the half-Heusler alloys \textbeta{}-LiMnZ (Z = N, P and Si) were
systematically examined for their electric, magnetic and stability
properties at optimized lattice constants and strained lattice constants
that exhibit half-metallic properties. Other phases of the half-Heusler
structure (\textalpha{} and \textgamma{}) are also reported here, but they are unlikely to be
grown. The magnetic moments of these stable Li-based alloys are expected
to reach as high as 4 \textmu{}\textsubscript{B} per unit cell when Z = Si and 5
\textmu{}\textsubscript{B} per unit cell when Z = N and P, however the
antiferromagnetic spin configuration is energetically favored when Z is
a pnictogen. \textbeta{}-LiMnSi at a lattice constant 14\% larger than its
equilibrium lattice constant is a promising half-metal for spintronic
applications due to its large magnetic moment and vibrational stability.
The modified Slater--Pauling rule for these alloys is determined.
Finally, a plausible method for developing half-metallic Li\(_x\)MnZ at
equilibrium, by tuning \(x\), is investigated, but, unlike
tetragonalization, this type of alloying introduces local structural
changes that destroy the half-metallicity.
\end{abstract}
\maketitle
\section{Introduction}\label{introduction}

Ternary compounds involving the Li atom in the form of half-Heusler, or
semi-Heusler, alloys, have recently attracted attention because of their
potential in optoelectronic and spintronic
applications\citep{Kieven:2010ef, Jungwirth:2011ej}. The crystal
structure, C1\textsubscript{b}, of any half-Heusler alloy is similar to
the structure, L2\textsubscript{1}, of a full-Heusler alloy
(X\textsubscript{2}YZ) but missing one X. Due to the missing element,
these alloys have three distinct atomic arrangements, called \textalpha{}-, \textbeta{}- and
\textgamma{}-phases\citep{Larson:2000ky}, used by various research groups
\citep{Kieven:2010ef, Jungwirth:2011ej, Larson:2000ky, Mancoff:1999kg, PhysRevLett.50.2024, Galanakis:2006iy, Toboia:2000gi}.
In \autoref{tbl:positions}, the positions occupied by the three atoms
and the vacancy are given according to the notations defined by
\citet{wyckoff1963crystal}. One of the authors (C.
F.)\citep{Kieven:2010ef} examined lithiated half-Heusler alloys in the
\textbeta{}-phase, namely LiMgZ (Z = N, P, As, Bi), LiYP, LiYP (Y = Zn, Cd) and
LiAlSi. They found that covalent bonding between the Y and Z atoms forms
the gap of these compounds. They suggested these half-Heusler alloys
could be used in optoelectronic and solar applications because the
values of the band gaps and lattice constants of these materials are
suitable for substituting CdS as buffer layer materials. Since there is
no magnetic element in any of these alloys, magnetic properties were not
addressed. \citet{Jungwirth:2011ej} considered a magnetic element in a
Li-based half-Heusler alloy and epitaxially grew \textbeta{}-LiMnAs on an InAs
substrate. They theoretically predicted that \textbeta{}-LiMnAs is
antiferromagnetic and confirmed it experimentally using a SQUID to
measure the magnetic moment.

\begin{table}
\caption{The positions of the three atoms, X, Y and Z in terms of the
Wyckoff notation: 4a = (0,0,0)\(a\), 4b = (1/2,1/2,1/2)\(a\), and 4c =
(1/4,1/4,1/4)\(a\), where \(a\) is the lattice parameter. The zinc
blende structure, YZ, is included for comparison. \label{tbl:positions}}
\begin{ruledtabular}
\begin{tabular}{@{}cccc@{}}
Structure & X & Y & Z\\
\midrule
\textalpha{} & 4c & 4b & 4a\\
\textbeta{} & 4b & 4a & 4c\\
\textgamma{} & 4a & 4c & 4b\\
Zinc blende & -- & 4a & 4c\\
\end{tabular}
\end{ruledtabular}
\end{table}

The studies of transition metal element (TME)-based half-Heusler alloys
indicate that there are a number of differences among the three phases.
Using the \textalpha{}-phase for TME-based half-Heusler alloys (X=Fe, Co, Ni; Y=Ti,
V, Zr, Nb, Mn), \citet{Toboia:2000gi} and \citet{Galanakis:2006iy}
showed that the d--d bonding between X and Y, of the half-metal (HM)
CrMnSb, is responsible for the states at the gap in the insulating
channel. A HM exhibits metallic properties in one spin channel and
semiconducting properties in the oppositely oriented spin channel. They
did not comment on the d--p bonding states between X and Z. We
previously suggested\citep{Shaughnessy:uku55sOX} a point of view in
Heusler alloys unifying the d--p and the d--d bonding: Z is the most
electronegative among the three atoms so it should form d--p bonding and
determine the primary bonding properties of the bonding--antibonding
gap. Additionally, the weak bonds between the two TM atoms, formed by
d--d bonding, contribute to the states at the gap. From this point of
view, the TME-based half-Heusler alloys, such as CrMnSb, in \textalpha{}- and
\textgamma{}-phases, differ by which of the two TM atoms is the nearest neighbor
(nn) of Z. Their properties differ accordingly.

How does our suggestion, \emph{to consider the nn of Z}, work for
Li-related half-Heusler alloys? We expect that the \textbeta{}-phase is
energetically favored because Z, the most electronegative element, is nn
to both Li and Mn. We also anticipate a larger gap than the TME-based
half-Heusler alloys since there is no d--d bonding forming states in the
gap. Based on the fact that Li easily gives up its outermost electron to
its neighbor, \textbeta{}-LiMnZ can be viewed as an intercalation of zinc blende
(MnZ)\(^-\) and Li\(^+\)\citep{Kieven:2010ef}. According the ionic
model\citep{Schwarz:1986p3411}, Mn loses three valence electrons to the
pnictogen Z when they are nn pairs, and the four remaining d-electrons
combine with the electron from Li to give a local magnetic moment as
large as 5 \textmu{}\textsubscript{B} per Mn. If Z is a group IV element, then
the maximum is 4 \textmu{}\textsubscript{B} per Mn.

Typical TME-based zinc blende alloys, such as MnZ, are not half-metallic
at their equilibrium lattice constant. The standard method to obtain
half-metallic properties from these alloys is to increase, or shrink,
their lattice constant by growing them as thin films on substrates with
larger, or smaller, lattice constants\citep{JJAP.39.L1118}. Straining
the lattice constant of the alloy can manifest half-metallic properties
by changing the bonding and exchange strength between
atoms\citep{Pask:2003fo}. These changes can shift the Fermi energy into
a gap of one spin channel while the Fermi energy cuts through states in
the other spin channel, resulting in a HM. We examine the alloys for
large magnetic moments or even half-metallic properties at lattice
constants away from equilibrium.

Since Li-based alloys, other than the ones included in this report, may
be desirable to grow, we determined the modified Slater--Pauling (SP)
rule\citep{Kubler:1984uj} for Li-based half-Heusler alloys based on the
band structure calculations. The rule provides a zeroth order
approximation to the magnetic moment of ferromagnets based on the number
of occupied states in the minority spin channel N\textsubscript{\ensuremath{\downarrow}}.
Combined with the fact that the total number of valence states is
\(N = N_{\downarrow}+N_{\uparrow}\), the magnetic moment is

\begin{align}
M &= N_{\downarrow}-N_{\uparrow}
  = N - 2N_{\downarrow}\label{eqn:sprule}
\end{align}

\noindent in units of \textmu{}\textsubscript{B}. The modified SP rule is useful
for predicting the magnetic moment of materials where N\textsubscript{\ensuremath{\downarrow}}
may remain constant, but N changes by replacing atoms with neighboring
atoms on the periodic table. The As mentioned previously,
\citet{Jungwirth:2011ej} already determined that \textbeta{}-LiMnAs favors the
antiferromagnetic configuration, so we examined the possibility that
ferromagnetism is energetically favored over antiferromagnetism in any
of the alloys.

For these alloys to be useful in devices, the issue of stability should
be addressed. In the past, the conventional wisdom is that the thin film
forms of HMs with lattice constants away from the equilibrium values are
stable up to \textasciitilde{}100 layers\citep{JJAP.39.L1118}, but we
question this claim. We compared the stability of \textbeta{}-LiMnP to meta-stable
ZB MnP, by considering phonon spectra determined by the response
function method\citep{Baroni:2001tn}.

With the Li-based alloys, it seems plausible to shift the Fermi energy
\(E_{\mathrm{F}}\) by adjusting the concentration of Li atoms. For an
alloy where its \(E_{\mathrm{F}}\) cuts through the bottom of the
conduction band of one spin channel, will lowering the concentration of
Li push the Fermi energy into the gap and give a HM at equilibrium? We
use the location of \(E_{\mathrm{F}}\) with respect to the gap as a
guide to investigate the possibility of half-metallicity in Li\(_x\)MnZ
alloys, with \(x < 1\).

In this paper, we investigate three LiMnZ with Z = N, P and Si and
address the following issues:

\begin{itemize}
\item
  What are the basic bonding and magnetic properties of LiMnZ,
  particularly in the lowest energy phase?
\item
  Compared to zinc bledne MnZ, what is the role of Li in the electronic
  and magnetic properties of LiMnZ?
\item
  Can any of the alloys energetically favor the ferromagnetic phase and
  give a large magnetic moment?
\item
  How does the phonon stability of \textbeta{}-LiMnZ compare to ZB MnZ and what
  are the implications for growth?
\item
  Finally, can adjusting the concentration of Li in
  \textbeta{}-Li\textsubscript{x}MnZ result in a HM at equilibrium?
\end{itemize}

In \autoref{structural-models-and-methods-of-calculation}, we discuss
the models used to answer these questions, and present a brief
description of the methods of calculation. Results and discussion of the
above issues will be presented in \autoref{results-and-discussion}.
Finally, in \autoref{summary}, we summarize our findings.

\section{Structural models and methods of
calculation}\label{structural-models-and-methods-of-calculation}

We used the primitive cell to find the equilibrium lattice constants by
means of minimizing the total energy of each alloy with respect to the
lattice constant. The phonon calculations also rely on the respective
primitive cells of \textbeta{}-LiMnP and ZB MnP. We calculate the
antiferromagnetic properties by constructing an tetragonal cell,
consisting of two formula units of LiMnZ, to allow the possibility of
antiparallel alignment of Mn atoms. Additionally, we explored the
possibility of lower concentrations of Li atoms by constructing
conventional unit cells consisting of four formula units of LiMnZ and
then removing one Li atom and letting the atoms and unit cell relax to
equilibrium. The resulting alloys are denoted
Li\textsubscript{3}Mn\textsubscript{4}Z\textsubscript{4}, or
Li\textsubscript{0.75}MnZ.

We used the spin-polarized version of the VASP
code\citep{Kresse:1996vf, Kresse:1996vk, Kresse:1994us, Kresse:1993ty},
which is based on density functional theory (DFT)\citep{hohenberg:B864}.
The generalized gradient approximation (GGA) of
\citet{PhysRevLett.77.3865} (PBE) was used to treat the
exchange-correlation between electrons. GGA provides realistic bonding
and magnetic properties, except for the value of the semiconducting gap.
The value of the semiconducting gap is not a crucial issue at this
point, so we did not consider more complicated methods utilizing
many-body techniques. Many-body methods improve upon the conduction
states in semiconductors and HMs\citep{Damewood:2011bi}, but also show
that half-metallic properties may disappear at finite
temperatures\citep{Chioncel:2006cl}.

The VASP package provides projector-augmented-wave (PAW)
pseudopotentials\citep{Blochl:1994dx} for Li, Mn, N, P, and Si that were
constructed using PBE. We used a basis of plane waves with a 1000 eV
kinetic energy cutoff for all calculations. The
Monkhorst---Pack\citep{monkhorst:5188} meshes of (17,17,17), (15,15,15)
and (11,11,11) were adopted for the Brillouin Zone (BZ) of the
primitive, tetragonal and the conventional cells, respectively. Using
these values, the convergence of the total energy and the magnetic
moment of any sample are better than 1.0 meV and 1.0
m\textmu{}\textsubscript{B}, respectively.

To address the stability of the lithiated compounds, we used the ABINIT
software package to perform response-function phonon
calculations\citep{Gonze:2009jb, Gonze:2005p1769}. We used the same
exchange-correlation functional as in VASP and used comparable
convergence parameters to calculate the ground state structure. To
calculate the force constants, a 4x4x4 irreducible \(\vec q\)-point
grid, centered at \(\vec q=(0, 0, 0)\) was constructed.

\section{Results and discussion}\label{results-and-discussion}

\subsection{Basic Properties of LiMnZ}\label{basic-properties-of-limnz}

Using the primitive unit cell, we first determined the equilibrium
lattice constants of the three compounds in the three phases (\textalpha{}-, \textbeta{}- and
\textgamma{}-phases). The results are summarized in \autoref{tbl:equilibrium}. The
lattice constants correlate with the covalent radii of Z
(r\textsubscript{cov}(N) = 0.71 \AA, r\textsubscript{cov}(P) = 1.09 \AA, and
r\textsubscript{cov}(Si) = 1.14 \AA\citep{Haynes:vx}. The \textbeta{}-phase is
consistently the lowest energy phase because Z, the most electronegative
element, is nn to both Mn and Li.

\begin{table}
\caption{The equilibrium lattice constants, total energies, and magnetic
moments for three lithiated half-Heusler alloys.
\label{tbl:equilibrium}}
\begin{ruledtabular}
\begin{tabular}{@{}cccc@{}}
Compound & \parbox[t]{60pt}{Lattice constant (\AA)} &
\parbox[t]{60pt}{Total energy relative to \textbeta{} (eV)} &
\parbox[t]{60pt}{Magnetic moment (\textmu{}\textsubscript{B}/Mn)}\\
\midrule
\textalpha{}-LiMnN & 4.961 & 2.379 & 4.675\\
\textbeta{}-LiMnN & 4.912 & 0.000 & 3.925\\
\textgamma{}-LiMnN & 5.139 & 1.992 & 4.805\\
\textalpha{}-LiMnP & 5.600 & 1.218 & 4.391\\
\textbeta{}-LiMnP & 5.717 & 0.000 & 4.090\\
\textgamma{}-LiMnP & 5.715 & 0.661 & 3.987\\
\textalpha{}-LiMnSi & 5.629 & 0.721 & 3.750\\
\textbeta{}-LiMnSi & 5.778 & 0.000 & 3.314\\
\textgamma{}-LiMnSi & 5.788 & 0.395 & 3.301\\
\end{tabular}
\end{ruledtabular}
\end{table}

\begin{figure}[htbp]
\centering
\includegraphics{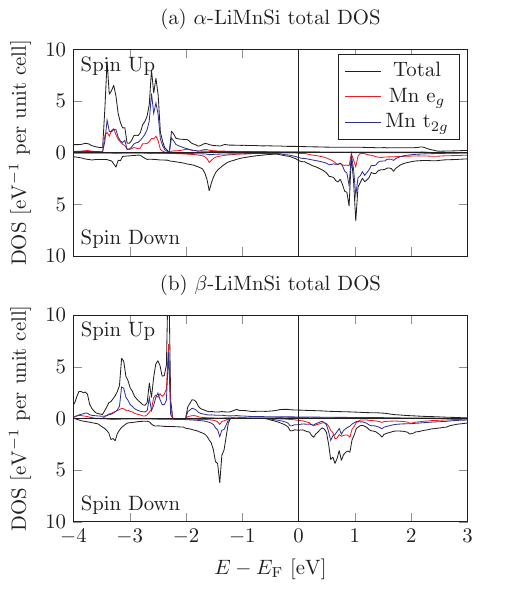}
\caption{Spin polarized total density of states (DOS) for LiMnSi in (a)
\textalpha{}- and (b) \textbeta{}- phases at their respective equilibrium lattice constants.
The Mn e\textsubscript{g} and t\textsubscript{2g} partial DOS are shown
as the shaded areas. \label{fig:alpha-and-beta-LiMnSi-opt}}
\end{figure}

In \autoref{fig:alpha-and-beta-LiMnSi-opt}, we show the density of
states (DOS) for \textalpha{}-and \textbeta{}-LiMnSi with partial DOS of the so-called
e\textsubscript{g} (doubly degenerate d\textsubscript{z2} and
d\textsubscript{x2-y2}) and t\textsubscript{2g} (triply degenerate
d\textsubscript{xy}, d\textsubscript{yz}, and d\textsubscript{zx})
states. Here, we choose to only discuss the bonding properties of LiMnSi
since, at the equilibrium lattice constant, the overall bonding features
are not significantly altered when Z = N or P. The \textbeta{}- and \textgamma{}- phase DOSs
are very similar---both have Mn and Si as nn and the position of the Li
atom does not drastically alter the DOS---so the \textgamma{}-phase is not
included. The \textbeta{}- and \textgamma{}- phases form a large bonding--antibonding gap,
due to the overlap of the t\textsubscript{2g} states of Mn and Si
sp\textsuperscript{3} states in the tetrahedral environment. The \textalpha{}-phase
is significantly different compared to the \textbeta{}- and \textgamma{}-phases and does not
form a large bonding--antibonding gap because the Mn and Si are second
neighbors in a cubic environment.

\subsection{The role of Li in LiMnZ}\label{the-role-of-li-in-limnz}

\begin{figure}[htbp]
\centering
\includegraphics{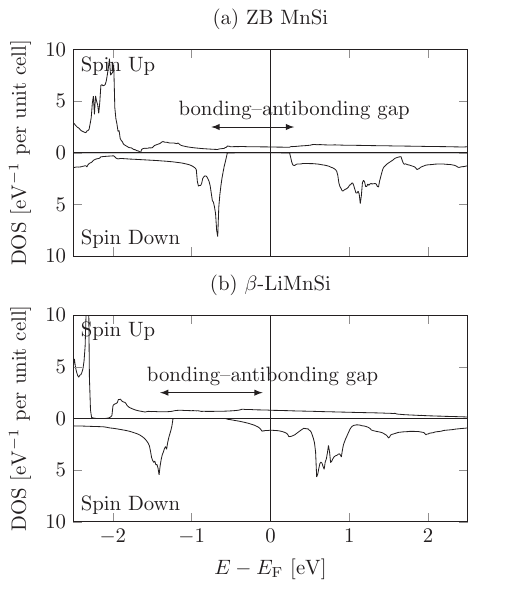}
\caption{Density of states for (a) zinc blende MnSi and (b) \textbeta{}-LiMnSi at
the optimized lattice constant. Arrow indicates the size of the
bonding--antibonding gap at the \textGamma{}
point.\label{fig:MnSi-beta-LiMnSi-opt}}
\end{figure}

\begin{figure}[htbp]
\centering
\includegraphics{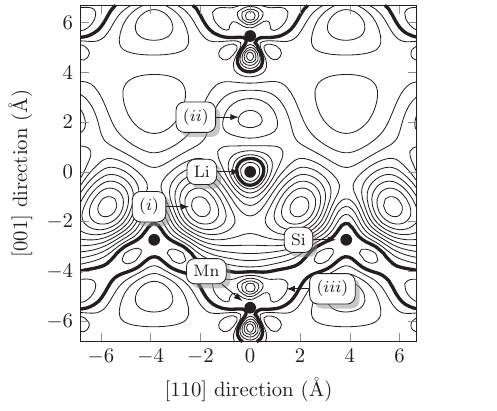}
\caption{Contour of the charge density difference (\textbeta{}-LiMnSi minus ZB
MnSi) at the same lattice constant. Circles indicate the location of
atoms. Labels (i) and (ii) indicates where the Li electron bonds with Si
and Mn, respectively, giving a positive contour. Label (iii) shows the
area where the bond between Mn and Si weakens and the contour region is
negative. The thick black lines indicate the zero contours.
\label{fig:den-diff-beta-LiMnSi-MnSi}}
\end{figure}

We have argued that Li easily gives up its valence electron to its
nearest neighbor. To substantiate this argument, we compare \textbeta{}-LiMnSi to
MnSi in the ZB structure. The latter is predicted to be a HM in which
the bonding, antibonding and and non-bonding states are easily
identified \citep{Pask:2003fo}. In \autoref{fig:MnSi-beta-LiMnSi-opt},
we compare the states around \(E_{\mathrm{F}}\) at \textGamma{} of the two spin
channels for \textbeta{}-LiMnSi and MnSi at the same lattice constant of 5.778 \AA,
and show the size of the bonding--antibonding gap (the energy difference
between the top of the valence band and the bottom of the conduction
band, in the minority spin channel, at the \textGamma{}-point). The primary effects
of Li are: (1) increasing the width of the occupied states and (2)
promoting its s-state to the p-states and contributing to the d--p mixed
states. The second effect indicates that Li gives up its electron. The
argument that the Li is giving up its electron is also reflected in the
charge density difference, \textbeta{}-LiMnSi minus MnP, shown in
\autoref{fig:den-diff-beta-LiMnSi-MnSi}. This plot shows the Li electron
redistributes its charge to Mn (i) and Si (ii), away from Li, and causes
the d--p hybridized bond between Mn and P to weaken, thereby shifting
the bond charge away from P and towards Mn (iii).

The pnictides in the \textbeta{}-phase do not have integer magnetic moments and
show no gap in either spin channel near \(E_{\mathrm{F}}\). According to
our criteria they are not HMs\citep{Fong:2008ii}. Due to the presence of
the highly electronegative pnictogen, the Li electron is not completely
transferred to the Mn. Instead, the electron from Li causes a reduction
in the strength of the d--p hybridization between the pnictogen and the
Mn under the tetrahedral environment. With the reduction in the
hybridization, the bonding--antibonding gap shrinks and the bottom of
the conduction band in the minority spin channel of \textbeta{}-LiMnZ becomes
occupied.

\subsection{Can the three half-Heusler alloys be ferromagnetic or
half-metallic?}\label{can-the-three-half-heusler-alloys-be-ferromagnetic-or-half-metallic}

We are interested in adjusting the lattice constant to force the values
of the magnetic moments to be as large as possible. In practice, the
consensus is that many devices are thin films and can be grown to match
the lattice constants on selected substrates. We can use our results to
predict the substrates that match the lattice constants of our compounds
with large magnetic moments, but we believe that this will not always be
possible for large lattice constants or energies far from equilibrium;
therefore, the issue of stability will be addressed later.
Tetragonalization can also occur during growth, but it will not
necessarily destroy half-metallicity\citep{Ozdogan:2012bq}. In
\autoref{tbl:hms}, we present detailed ferromagnetic properties of the
three compounds in the primitive cells of the three types of structures.

\begin{table}
\caption{Summary of the lattice constants, total energies, magnetic
moments per unit cell, and energy gaps near the Fermi energy
\(E_{\mathrm{F}}\) (with spin channel) of the three compounds in the
three phases where the magnetic moment is largest or integer.
\label{tbl:hms}}
\begin{ruledtabular}
\begin{tabular}{@{}lcccc@{}}
Compound & \parbox[t]{40pt}{Lattice constant (\AA)} &
\parbox[t]{60pt}{Total energy relative to equilibrium \textbeta{} (eV)} &
\parbox[t]{45pt}{Magnetic Moment (\textmu{}\textsubscript{B}/Mn)} &
\parbox[t]{45pt}{Energy gap near $E_{\mathrm{F}}$ (eV)}\\
\midrule
\textalpha{}-LiMnN & 5.300 & 0.274 & 5.000 & 0.819 \ensuremath{\downarrow}\\
\textbeta{}-LiMnN & 5.641 & 1.097 & 4.990 & Metal\\
\textgamma{}-LiMnN & 5.571 & 2.463 & 5.000 & Metal\\
\textalpha{}-LiMnP & 6.250 & 0.668 & 5.000 & 0.341 \ensuremath{\downarrow}\\
\textbeta{}-LiMnP & 6.550 & 0.944 & 5.000 & 1.693 \ensuremath{\downarrow}\\
\textgamma{}-LiMnP & 7.248 & 3.326 & 5.000 & Metal\\
\textalpha{}-LiMnSi & 6.274 & 1.384 & 4.050 & Metal\\
\textbeta{}-LiMnSi & 6.250 & 0.318 & 4.000 & 0.915 \ensuremath{\downarrow}\\
\textgamma{}-LiMnSi & 6.300 & 0.392 & 4.000 & 0.723 \ensuremath{\downarrow}\\
\end{tabular}
\end{ruledtabular}
\end{table}

Every alloy in \autoref{tbl:hms} with an integer magnetic moment and a
gap is a HM. These include: \textalpha{}-LiMnN, \textalpha{}-LiMnP, \textbeta{}-LiMnSi, \textbeta{}-LiMnP, and
\textgamma{}-LiMnSi. The lattice constant can be increased slightly without
destroying the half-metalicity since there is a range of lattice
constants where \(E_{\mathrm{F}}\) falls within the gap of the
semiconducting channel. The magnetic moments of the HMs agree with the
predictions of the ionic model plus the contribution of the Li electron
to the moment of Mn. The remaining alloys are ferromagnets and any
change in the lattice constant will decrease the moment. Some of the
important features of the individual half-Heusler alloys are discussed
below.

\begin{figure}[htbp]
\centering
\includegraphics{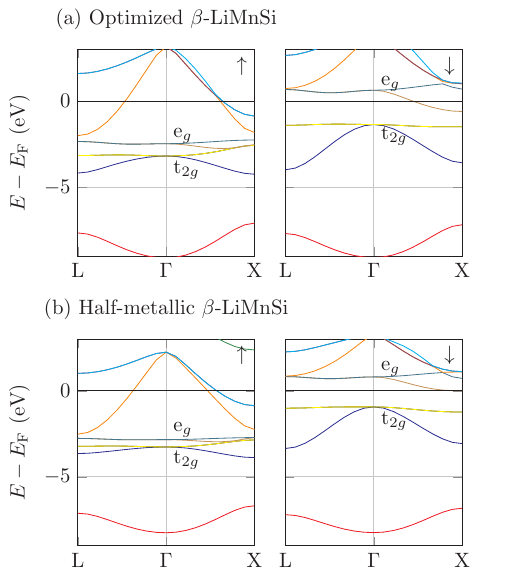}
\caption{Band structure for \textbeta{}-LiMnSi along L-\textGamma{}-X at (a) the equilibrium
lattice constant \(a = 5.778\) \AA, and (b) the half-metallic lattice
constant \(a = 6.250\) \AA. The triply degenerate states,
t\textsubscript{2g}, and the doubly degenerate states,
e\textsubscript{g}, near \(E_{\mathrm{F}}\) are labeled.
\label{fig:bands-LiMnSi-gamma}}
\end{figure}

In \autoref{fig:bands-LiMnSi-gamma}, we show the band structures of
\textbeta{}-LiMnSi at two lattice constants: the equilibrium lattice constant and
the half-metallic lattice constant. The lowest bands are from the
s-orbitals of Si. The next bands, shown in the figures, are the
so-called t\textsubscript{2g} states triply degenerate states at the
\textGamma{}-point that split to doubly (upper) and singly (lower) degenerate
states as \(\vec k\) moves towards X. The next higher energy states at
the \textGamma{}-point are originally from e\textsubscript{g} states of Mn. Since
their lobes point away from nn Si and toward the second neighbor (sn) of
the Mn, they do not strongly interact with any other states so they are
called the non-bonding states. The half-metallic gap, in
\autoref{fig:bands-LiMnSi-gamma}(b), is formed between the
e\textsubscript{g} states and the doubly degenerate states that split
off of the t\textsubscript{2g} states. Using the band structure, we can
understand why this half-Heusler alloy is not a HM at the equilibrium
lattice constant. The e\textsubscript{g} states are insensitive to the
separation between nn, but the smaller equilibrium lattice constant
causes \(E_{\mathrm{F}}\) to shift up, and intersect with, the
e\textsubscript{g} states (\autoref{fig:bands-LiMnSi-gamma}(a))
resulting in the disappearance of the half-metalicity.

The reason that \textalpha{}-LiMnSi is not a HM can be attributed to the fact that
Si forms nn pair with Li and sn pair with Mn. The qualitative details
are: (i) Ideally, the Li electron can form a covalent bond with Si
because the electronegativity of Si is not as large as either P or N.
(ii) The sn configuration between Si and Mn favors the p-states of Si to
form bonds with three Mn electrons. The bonds are not strong enough to
form a gap. Consequently, the TM does not completely transfer its 3
d-electrons to Si and gives a moment of 4.050 \textmu{}\textsubscript{B}. The
pnictides exhibit half-metallic properties in the \textalpha{}-phase at lattice
constants larger than their respective equilibrium values. The gap for
\textalpha{}-LiMnP is 0.376 eV while it is 0.696 eV for \textalpha{}-LiMnN. The gaps reflect
the strength of the electronegativity of P with respect to N and the
differing lattice constants. The gaps are formed by the
t\textsubscript{2g} states of Mn and the sp\textsuperscript{3} states of
the pnictogen, as in the \textbeta{}- and \textgamma{}- phases, but with a much smaller
bonding--antibonding gap.

Next, we focus on the antiferromagnetic cases. In order to investigate
this phase, we use the tetragonal cell so that there are two Mn atoms
which can have their magnetic moments oriented oppositely. In
\autoref{tbl:antiferro}, we present the results for the three
half-Heusler alloys in the antiferromagnetic phase and their energy
difference with respect to the ferromagnetic half-metallic case. From
our calculations, all three phases of LiMnSi can be ferromagnetic HMs
while LiMnN and LiMnP can be antiferromagnetic HMs at the half-metallic
lattice constant, much like LiMnAs\citep{Jungwirth:2011ej}.

\begin{table}
\caption{Summary of lattice constant and the energy difference between
AFM and FM orderings of the three compounds at lattice constants given
in \autoref{tbl:hms}. Negative energy differences indicate that the AFM
phase is energetically preferred. \label{tbl:antiferro}}
\begin{ruledtabular}
\begin{tabular}{@{}cc@{}}
Compound & E\textsubscript{AFM}-E\textsubscript{FM} Relative Total
Energy (eV)\\
\midrule
\textalpha{}-Li\textsubscript{2}Mn\textsubscript{2}N\textsubscript{2} & -0.037\\
\textbeta{}-Li\textsubscript{2}Mn\textsubscript{2}N\textsubscript{2} & -0.292\\
\textgamma{}-Li\textsubscript{2}Mn\textsubscript{2}N\textsubscript{2} & -0.302\\
\textalpha{}-Li\textsubscript{2}Mn\textsubscript{2}P\textsubscript{2} & -0.026\\
\textbeta{}-Li\textsubscript{2}Mn\textsubscript{2}P\textsubscript{2} & -0.123\\
\textgamma{}-Li\textsubscript{2}Mn\textsubscript{2}P\textsubscript{2} & -0.087\\
\textalpha{}-Li\textsubscript{2}Mn\textsubscript{2}Si\textsubscript{2} & +0.100\\
\textbeta{}-Li\textsubscript{2}Mn\textsubscript{2}Si\textsubscript{2} & +0.347\\
\textgamma{}-Li\textsubscript{2}Mn\textsubscript{2}Si\textsubscript{2} & +0.324\\
\end{tabular}
\end{ruledtabular}
\end{table}

\subsection{Can the lithiated alloys in the \textbeta{}-phase be more stable than
the corresponding compounds in the zinc-blende
structure?}\label{can-the-lithiated-alloys-in-the-ux3b2-phase-be-more-stable-than-the-corresponding-compounds-in-the-zinc-blende-structure}

\begin{figure}[htbp]
\centering
\includegraphics{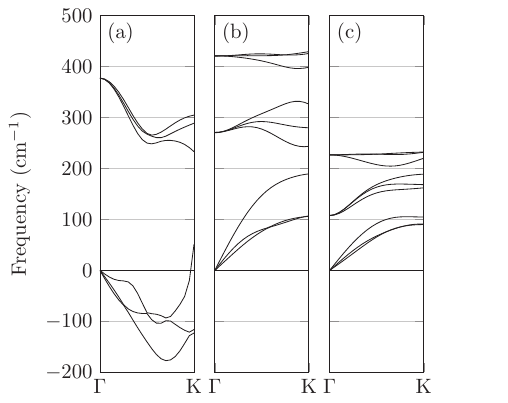}
\caption{The phonon bands along \textGamma{}--K of (a) zinc blende MnSi, (b)
\textbeta{}-LiMnSi at its equilibrium lattice constant \(a = 5.778\) \AA and (c)
\textbeta{}-LiMnSi at the lattice constant in which it is half-metallic:
\(a = 6.590\) \AA. Negative values indicate imaginary (unstable)
frequencies.\label{fig:bands-phonon}}
\end{figure}

We carried out response-function phonon calculations on a 4x4x4
\(\vec q\)-point grid for ZB MnSi at its equilibrium lattice constant
and \textbeta{}-LiMnSi at two lattice constants: its equilibrium value and the
lattice constant where the compound is half-metallic. The results are
shown in \autoref{fig:bands-phonon}. MnZ in the zinc-blende structure
has an unstable transverse acoustic (TA) branch along the zone boundary
in the {[}110{]} direction (\textGamma{}--K). This agrees with the known fact that
the ZB structure for MnP-type compounds is not the ground state of these
compounds\citep{JJAP.39.L1118}. \textbeta{}-LiMnSi, at its equilibrium lattice
constant, shows stability in the {[}110{]} direction. At the lattice
constant that gives half-metallic properties, the six optical branches
of \textbeta{}-LiMnSi are also stable. The addition of the Li atom in the
structure increases the restoring force against sheer stress.

\begin{figure}[htbp]
\centering
\includegraphics{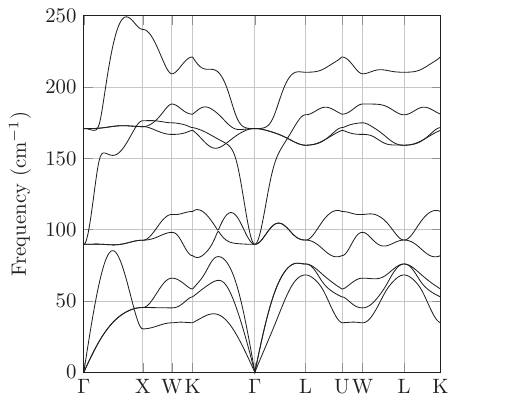}
\caption{The full-zone phonon bands of \textbeta{}-LiMnSi at a lattice constant
showing half-metallic properties: \(a = 6.590\) \AA.
\label{fig:bands_hm_fz}}
\end{figure}

The response-function method allows for the determination of the phonon
bands in the full Brillouin zone. The full-zone phonon bands for
\textbeta{}-LiMnSi at the half-metallic lattice constant are provided in
\autoref{fig:bands_hm_fz}

\subsection{Li\textsubscript{x}MnZ with x \textless{}
1}\label{lixmnz-with-x-1}

\begin{figure}[htbp]
\centering
\includegraphics{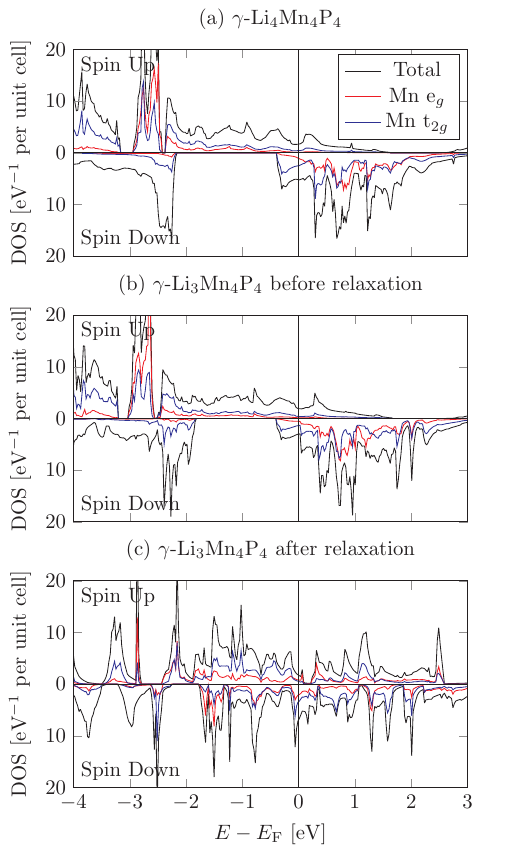}
\caption{DOS for (a)
Li\textsubscript{4}Mn\textsubscript{4}P\textsubscript{4}, (b)
Li\textsubscript{3}Mn\textsubscript{4}P\textsubscript{4} before
relaxation and (c)
Li\textsubscript{3}Mn\textsubscript{4}P\textsubscript{4} after cell and
ion relaxation. Shaded regions show the partial DOS for the
e\textsubscript{g} and t\textsubscript{2g} states around the Mn
atoms.\label{fig:limnp-gamma-pdos}}
\end{figure}

Based on the fact that the density of states of \textgamma{}-LiMnP, given in
\autoref{fig:limnp-gamma-pdos}(a), shows \(E_{\mathrm{F}}\) slightly
above the conduction band edge in the minority spin channel, we decided
to find new HMs by removing one Li from the conventional cell. The
symmetry of the half-Heusler phases allows the removal of any Li atom in
the conventional unit cell. The resultant sample is labeled as
\textgamma{}-Li\textsubscript{3}Mn\textsubscript{4}P\textsubscript{4} in the
conventional cell or \textgamma{}-Li\textsubscript{0.75}MnP as an alloy. We
anticipated that the removal of a Li atom could lower \(E_{\mathrm{F}}\)
into the gap and produce a HM. However, since the Li gave up its
electron to other states, an imbalance in the forces drove the atoms,
nearest to the vacancy, toward the vacancy and they formed bonds. The
resulting compound is not a HM. The bond length between Mn atoms was
4.041 \AA before relaxation and 2.452 \AA after relaxation. The optimized
lattice constant after relaxation is 5.418 \AA. The confinement of the
spins causes them to anti-align and results in a lower magnetic moment,
similar to what we previously found with MnC\citep{qian:7459}. The
magnetic moment of
\textgamma{}-Li\textsubscript{3}Mn\textsubscript{4}P\textsubscript{4} reduces to
1.145 \textmu{}\textsubscript{B} per Mn atom.

\subsection{Mofified Slater--Pauling
Rule}\label{mofified-slaterpauling-rule}

The lithiated half-Heusler alloys either have 13 or 12 valence electrons
per unit cell. Since the total number of valence electrons \(N\) is
smaller in these alloys than in the TM-related compounds, the modified
SP rule should definitely be modified. The band structure of \textbeta{}-LiMnSi in
the primitive cell, shown in \autoref{fig:bands-LiMnSi-gamma}(b), shows
four bands in the semiconducting channel, so \(N_\downarrow = 4\). From
\autoref{eqn:sprule}, the modified SP rule is

\begin{align}
M = (N - 8)\text{\textmu{}\textsubscript{B}},\label{eqn:Slater-Pauling}
\end{align}

\noindent so for LiMnSi, \(N = 12\) and M is 4 \textmu{}\textsubscript{B} per
formula unit. Similarly, M is 5 \textmu{}\textsubscript{B} per formula unit for
the pnictides. The calculated moments in \autoref{tbl:hms} agree well to
the predicted results of \autoref{eqn:Slater-Pauling}, however the
modified SP rule is not able to account for the antiferromagnetism in
the cases where Z is a pnictogen.

\section{Summary}\label{summary}

The fact that Li easily gives up its electron to Mn led us to
investigate three Li-based half-Heusler alloys, each involving one TM
element per formula, for the possibility of them being ferromagnetic
with large magnetic moments. Three different arrangements of the atoms
inside a unit cell, denoted \textalpha{}-, \textbeta{}- and \textgamma{}-phases, appear in the
literature. We offer a unified view of bonding in the three alloys based
on previous studies on TM-related half-Heulser alloys involving two
TMEs\citep{Shaughnessy:uku55sOX}. The strongest bond is formed between
the Z and Mn resulting in the primary bonding--antibonding gap. To
understand the role played by the Li, we compared the bonding properties
of \textbeta{}-LiMnSi, a prototype, to MnSi, a HM in zinc blende structure. Using
the tetragonal cell, we found that \textalpha{}-LiMnP, \textalpha{}-LiMnN and \textbeta{}-LiMnP are
antiferromagnetic while LiMnSi is a ferromagnetic HM in the \textbeta{} and \textgamma{}
phases. The reasons for the above facts are provided. The magnetic
moments for the ferromagnetic alloys are all larger than 3
\textmu{}\textsubscript{B} per formula unit. These alloys should be good
candidates as spintronic materials for devices operating at or above
room temperature. A new modified SP rule predicting the magnetic moments
in this class of half-Heusler alloys is proposed and predicts well for
the three alloys with half-metallic properties, however the rule is
incapable of predicting antiferromagnetism. The stability against shear
stress, as compared to the simple ZB structure, is demonstrated by
calculating the phonon spectrum, {[}110{]} direction of the Brillouin
zone, using the response function scheme. We show that MnP is unstable
along the {[}110{]} direction and find that the half metallic Li-based
alloys can be stable. Finally, we show that the removal of Li atoms will
not lower \(E_{\mathrm{F}}\) into the gap region as we originally
believed. Instead, the atoms surrounding the vacancy left by the Li will
relax towards the opening, bond, and then destroy the half-metallicity.
Hopefully, these results will facilitate the search and subsequent
growth of new HMs involving alkali or even alkaline metals.

\bibliography{paper}

\begin{acknowledgments}
Work at UC Davis was supported in part by the National Science
Foundation Grant No. ECCS-0725902. Work at Lawrence Livermore National
Laboratory was performed under the auspices of the U.S. Department of
Energy by Lawrence Livermore National Laboratory under Contract
DE-AC52-07NA27344. One of the authors (L. D.) would like to thank Barry
Klein for useful discussions."
\end{acknowledgments}
\end{document}